\documentclass[conference, comsoc]{IEEEtran}
\IEEEoverridecommandlockouts
\usepackage[left=0.625in,right=0.625in,top=0.76in,bottom=1.1in]{geometry}

\usepackage{subcaption}
\usepackage{multirow}
\usepackage{array}
\usepackage{balance}
\usepackage{cite}
\usepackage{amsmath,amssymb,amsfonts}
\usepackage{graphicx}
\usepackage{textcomp}
\usepackage{xcolor}
\usepackage{bm}
\usepackage{nicefrac}
\usepackage{mathtools}
\usepackage{comment}
\usepackage{pgfplots}
\usepackage{pgfkeys,pgfmath,pgfcore}
\pgfplotsset{compat=1.18}
\pgfkeys{/pgf/number format/.cd,fixed,precision=2}
\usepackage[ruled,vlined]{algorithm2e}
\usepackage{multirow}
\usepackage[colorlinks=true,allcolors=blue,bookmarks=false]{hyperref}
\def\BibTeX{{\rm B\kern-.05em{\sc i\kern-.025em b}\kern-.08em
    T\kern-.1667em\lower.7ex\hbox{E}\kern-.125emX}}

\DeclareMathOperator{\myvec}{vec}
\begin{document}




\title{Self-Supervised Radio Pre-training: Toward \\ Foundational Models for Spectrogram Learning}



\author{\IEEEauthorblockN{ Ahmed Aboulfotouh\IEEEauthorrefmark{3},
Ashkan Eshaghbeigi\IEEEauthorrefmark{1},
Dimitrios Karslidis\IEEEauthorrefmark{1}, and 
Hatem Abou-Zeid\IEEEauthorrefmark{3}
}
\IEEEauthorblockA{\IEEEauthorrefmark{3}{Department of Electrical and Software Engineering}, 
{University of Calgary}, Canada}
\IEEEauthorblockA{\IEEEauthorrefmark{1}{Qoherent Inc.}, 
{Toronto, Ontario, Canada}} 
\thanks{The authors would like to thank Qoherent Inc. and  MITACS Accelerate for their support of this research. The authors would also like to thank Denvr Dataworks, Calgary, Canada for their high-performance compute used to conduct this research.}
}

\maketitle




\begin{abstract}

Foundational deep learning (DL) models are general models, trained on large, diverse, and unlabelled datasets, typically using self-supervised learning techniques have led to significant advancements especially in natural language processing. 
These pretrained models can be fine-tuned for related downstream tasks, offering faster development and reduced training costs, while often achieving improved performance. 
In this work, we introduce Masked Spectrogram Modeling, a novel self-supervised learning approach for pretraining foundational DL models on radio signals. Adopting a Convolutional LSTM architecture for efficient spatio-temporal processing, we pretrain the model with an unlabelled radio dataset collected from over-the-air measurements. Subsequently, the pretrained model is fine-tuned for two downstream tasks: spectrum forecasting and segmentation. Experimental results demonstrate that our methodology achieves competitive performance in both forecasting accuracy and segmentation, validating its effectiveness for developing foundational radio models.

\end{abstract}


\section{Introduction}
\label{sec:intro}

A foundational model is a general model pretrained on a large-scale - usually unlabeled - dataset, typically through self-supervised learning \cite{ssl_survey}. Through this training, the model develops a solid understanding of the target modality, such as text in natural language processing (NLP) or images in computer vision. This understanding allows the model to be fine-tuned for diverse downstream tasks. Foundational models in NLP \cite{bert, roberta} and computer vision \cite{vis_in_wild} have driven significant advancements through leveraging the knowledge encoded in their pretrained representations. This facilitates quicker experimentation, more efficient resource utilization, and potentially, improved performance on downstream tasks that smaller models or those with more limited domain knowledge cannot achieve.

Deep learning has showcased promising results when applied in wireless communication \cite{dl_comm_1}. The effectiveness has been demonstrated across various tasks, including automatic modulation classification \cite{amc_1}, channel estimation \cite{chan_estim_1}, constellation and waveform design \cite{waveform_1}, among others. However, these models are highly specialized, echoing the early stages of deep learning's evolution in NLP and computer vision. The reliability of these models across data distribution shifts and their ability to generalize is also usually limited.

Introducing the concept of foundational models into wireless communication holds substantial promise to overcome these limitations \cite{llm_telecom}.
We argue that as in NLP and computer vision, where a wealth of unlabeled data exists —  communication signals can be harnessed for pretraining such foundational models through self-supervised learning, mitigating the expense associated with data labeling. 
Moreover, leveraging a foundational model as a backbone for multiple downstream tasks, which utilize its pretrained representations in subsequent processing, reduces computational demands. This approach can also improve generalization by leveraging the broader knowledge encoded within foundational model representations compared to highly specialized models which suffer from limited scope.

Drawing inspiration from these advancements, particularly in \cite{bert, maskedaudio, vis_in_wild}, we introduce a foundational radio model pretrained using masked spectrogram modelling (MSM) — a novel technique, we propose for wireless signals. This model is then fine-tuned to perform two different downstream tasks: spectrogram forecasting, which involves predicting future spectrogram based on past data, and spectrogram segmentation, which consists of distinguishing between background noise and other signal activities within the spectrogram. These tasks, while different, are complementary in the context of spectrum analysis and constitute a usage scenario for a foundational model integrated in a opportunistic spectrum access system. 
The primary contributions of our paper are:
\begin{itemize}
    \item We propose and develop a novel self-supervised learning approach, MSM, for pre-training foundational models on radio signals. To the best of our knowledge, this work represents the first demonstration of radio foundational models for spectrogram learning using unlabeled data. 
    \item We demonstrate the effectiveness of the proposed approach utilizing a real-world dataset that we collected over a software-defined radio testbed. The recordings are time-domain IQ samples received between 2.4 to 2.65 GHz.
    \item  Our results show that the developed MSM approach is able to learn features that generalize to both related and unrelated downstream tasks. Fine-tuning the foundational model demonstrated competitive results for spectrum forecasting and spectrum segmentation which had a distinctly different and unseen data distribution. 
\end{itemize}

The results of this paper highlight the significant potential that radio foundational models have to effectively enable multiple downstream spectrogram tasks. It is envisioned that such models will foster wider adoption of AI to enable reliable network performance and services. 

The remainder of the paper is structured as follows: 
Section \ref{sec:datasets} presents the two datasets utilized for pretraining the foundational model, and for the spectrum forecasting and segmentation tasks.
Section \ref{sec:methods} outlines the architecture and algorithm of the self-supervised foundational model.
Section \ref{sec:results} presents numerical experiments conducted to evaluate the proposed methodology.
Finally, Section \ref{sec:conclusion} concludes the paper.

\section{Testbed and Datasets}
\label{sec:datasets}
We leverage two datasets in this paper. The first is a Real-time Radio Dataset (RRD), captured in real-time using a software-defined radio (SDR) test-bed developed with PlutoSDRs. The second dataset  simulates 5G New Radio (NR) and LTE transmissions in neighboring bands. This is called the Segmentation Dataset (SD). In both datasets, our primary emphasis is on processing spectrogram data rather than IQ samples, thus a significant portion of preprocessing and data preparation revolves around spectrogram computation.


\subsection{Real-time Radio Dataset (RRD)}
The RRD dataset consists of time-domain recordings of IQ samples, which represent both the in-phase (I) and quadrature (Q) components of the RF signal. Each recording corresponds to a distinct center frequency, sampling frequency, and running for a specific duration. The center frequency spans from $2.4$ to $2.65$ GHz, with the sampling frequency varying between $10$ MHz and $60$ MHz. The time duration typically averages around $100$ ms. The data was collected in downtown Toronto, Canada. We utilize the dataset for foundational model pretraining and spectrum forecasting. There are $240$ recordings in total corresponding to approximately $24$ seconds of RF activity.

\noindent \textbf{Spectrogram Computation.}
The spectrogram of each IQ recording is then computed as follows.
\begin{enumerate}
    \item Divide the recording into non-overlapping $2$ ms slices.
    \item Compute the spectrogram for each $2$ ms slice.
    \item Convert each spectrogram from the linear to the log scale.
\end{enumerate}

The parameters used to generate the RRD dataset are summarized in Table \ref{tab:rrd_params}.

\begin{table}[h!]
    \caption{RRD Dataset Generation Parameters}
    \renewcommand{\arraystretch}{1.5}
    \setlength{\tabcolsep}{12pt}
    \centering
    \begin{tabular}{|cc|c|}
    \hline
    \multicolumn{2}{|c|}{\textbf{Parameters}}  & \textbf{Value} \\ \hline
    
    \multicolumn{1}{|c|}{\multirow{4}{*}{\begin{tabular}[c]{@{}c@{}}\textbf{Spectrogram} \\ \textbf{Parameters}\end{tabular}}} & FFT Size  &  $1024$ \\ \cline{2-3} 
    \multicolumn{1}{|c|}{} & Window Function &  Hanning \\ \cline{2-3} 
    \multicolumn{1}{|c|}{} & Window Size & $512$ \\ \cline{2-3} 
    \multicolumn{1}{|c|}{} & Hop Size & $512$ \\ \hline
    
    \multicolumn{1}{|c|}{\multirow{3}{*}{\begin{tabular}[c]{@{}c@{}}\textbf{Slicing} \\ \textbf{Parameters}\end{tabular}}} & Sentence Duration  & $10$, $20$ ms \\ \cline{2-3} 
    \multicolumn{1}{|c|}{} & Sentence Shape  & $(256, 256)$ \\ 
    \cline{2-3} 
    \multicolumn{1}{|c|}{} & Token Shape  & $(256, 16)$ \\ \hline
    \end{tabular}
    \vspace{5pt}
    \label{tab:rrd_params}
\end{table}

\subsection{NR-LTE Segmentation Dataset (SD)}

The process of creating the SD dataset begins with the generation of 5G NR and LTE signals individually. Subsequently, these signals are transmitted through their respective wireless channels in adjacent bands. We employ the Matlab Communication Toolbox for signal generation, following the guidelines outlined in \cite{5g_segmentation_matlab}. The parameters for generating 5G NR and LTE signals are presented in Tables \ref{tab:5g_nr_params} and \ref{tab:lte_params} respectively.

\begin{table}[h!]
    \renewcommand{\arraystretch}{1.5}
    \caption{5G NR Signal Generation Parameters.}
    \setlength{\tabcolsep}{12pt}
    \centering 
    \begin{tabular}{|c|c|}
    \hline
    \textbf{Parameter} & \textbf{Value} \\
    \hline
    Bandwidth &   $10, 15, \cdots, 50$ MHz    \\
    \hline
    Sub-Carrier Spacing (SCS) & $15, 30$ KHz \\
    \hline
    \begin{tabular}[c]{@{}c@{}}Synch. Signal  Block (SSB) Pattern\end{tabular} & Cases A and B \\
    \hline
    \begin{tabular}[c]{@{}c@{}}Synch. Signal  Block (SSB) Period\end{tabular}  & $20$ ms\\
    \hline
    \end{tabular}
    \vspace{5pt}
    \label{tab:5g_nr_params}
\end{table}

\begin{table}[h!]
    \caption{LTE Signal Generation Parameters.}
    \renewcommand{\arraystretch}{1.5}
    \setlength{\tabcolsep}{12pt}
    \centering
    \begin{tabular}{|c|c|}
        \hline
        \textbf{Parameter} & \textbf{Value} \\
        \hline
         Bandwidth & $5, 10, 15, 20$ MHz \\
        \hline
         Reference Channel & R. \{2, 4, 6, 8\} \\
        \hline
         Duplex Mode & FDD \\
         \hline
    \end{tabular}
    \vspace{5pt}
    \label{tab:lte_params}
\end{table}


In more detail, the dataset creation process involves two main steps: signal generation and spectrogram computation. 

\noindent \textbf{Signal Generation}
\begin{enumerate}
    \item Randomly select a signal configuration from Table \ref{tab:5g_nr_params} for 5G NR and Table \ref{tab:lte_params} for LTE. Generate $40$ subframes of signal transmission, corresponding to $40$ ms.
    \item Apply the respective signal through its corresponding multipath fading channel. For 5G NR, the NR clustered delay line channel is utilized, while for LTE, the LTE fading channel is employed. 
    \item Perform frequency up-conversion on both signals to position them in neighboring bands, then mix the signals in time. Operate at a center frequency of $4$ GHz with a sampling rate of $61.44$ MHz. Randomly place the signals within the band-of-interest, ensuring no frequency overlap.
\end{enumerate}

\noindent \textbf{Spectrogram Computation} 
\begin{enumerate}
    \item Compute the spectrogram for the resulting signal mixture.
    \item Resize the spectrogram to the shape $(256, 256)$.
    \item Create a label image of shape $(256, 256)$, assigning a value of $1$ to pixels with NR signals, $2$ to pixels with LTE signals, and $0$ to pixels with noise.
    \item Store the spectrogram and label image pair. 
\end{enumerate}

\begin{figure}[t!]
    \centering
    \includegraphics[width=0.85\linewidth, keepaspectratio]{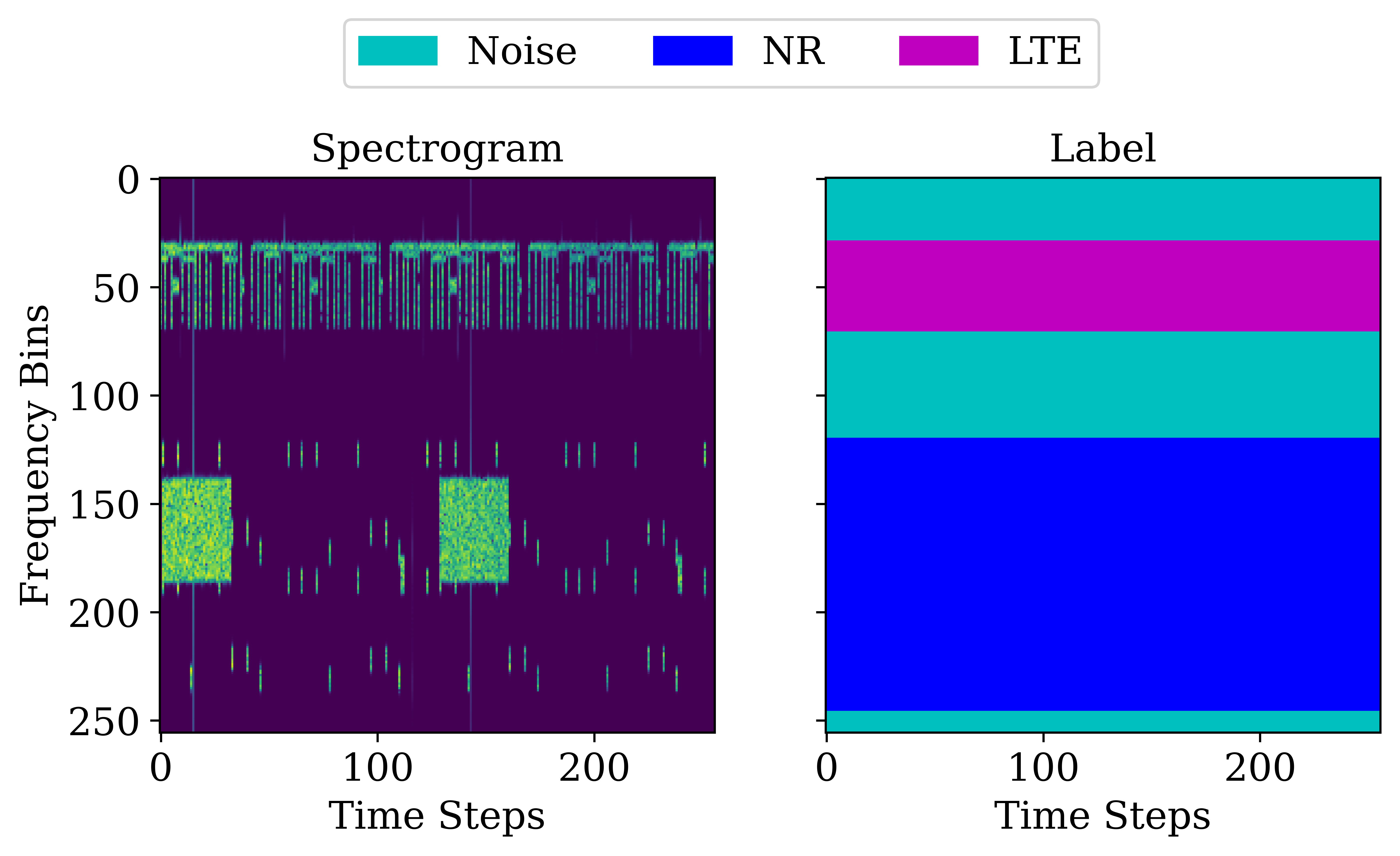}
    \caption{A spectrogram and label pair for the segmentation task.}
    \label{fig:spectrogram_label_segment}
\end{figure}

\noindent A sample spectrogram of this dataset is shown in Figure \ref{fig:spectrogram_label_segment}.

It is worth noting that the range of noise power is handled differently between the training and test sets. For the training set, we utilize a normal distribution $\mathcal{N}(-70, 5)$ dBm, while for the test set, we employ a uniform distribution $\mathcal{U}(-90, -20)$ dBm.
The reasoning behind this approach is as follows: in the training phase, we aim to prevent instances with high noise power from dominating the loss function, which could skew the training process. To achieve this, we undersample high noise instances by using a normal distribution. However, during testing, we want the model to be evaluated across all noise power levels equally. Therefore, we opt for a uniform distribution to ensure fair testing conditions.
\section{Foundational Model for Spectrogram Learning}
\label{sec:methods}

In this section we first present the methodology we propose to create the equivalent of \emph{sentences} and \emph{tokens} in the context of spectrograms. These radio sentences are then utilized by the proposed self-supervised masked spectrogram modelling approach which we present next. Here we utilize a convolutional LSTM (ConvLSTM) model introduced in \cite{convlstm}. This model is specifically crafted to capture crucial spatio-temporal features, aligning with our spectrogram learning needs. The convolutional component focuses on spatial properties, while the LSTM configuration handles temporal aspects. It accepts a sequence of two-dimensional spectrogram tokens as input and produces an output sequence of equal length. The details of the two downstream tasks that leverage this foundational ConvLSTM are then presented. 

\subsection{Creation of Radio Sentences and Tokens}
\begin{enumerate}
\item[(a)] Randomly sample a sequence of successive spectrograms with a duration ranging from $10$ to $20$ ms.
\item[(b)] Concatenate the sequence of spectrograms along the time-axis.
\item[(c)] Resize the result to a shape of $(256, 256)$.
\item[(d)] Divide the result along the time-axis into a sequence of tokens with a shape of $(256, 16)$, allowing the sequence of tokens to be represented as a 3D array with a shape of $(16 ,256, 16)$.
\item[(e)] Append the resulting sentence—a sequence of tokens—to the corpus, which will contain sentences of variable size.
\end{enumerate}

\begin{figure}[t!]
    \centering
    \includegraphics[width=0.6\linewidth, keepaspectratio]{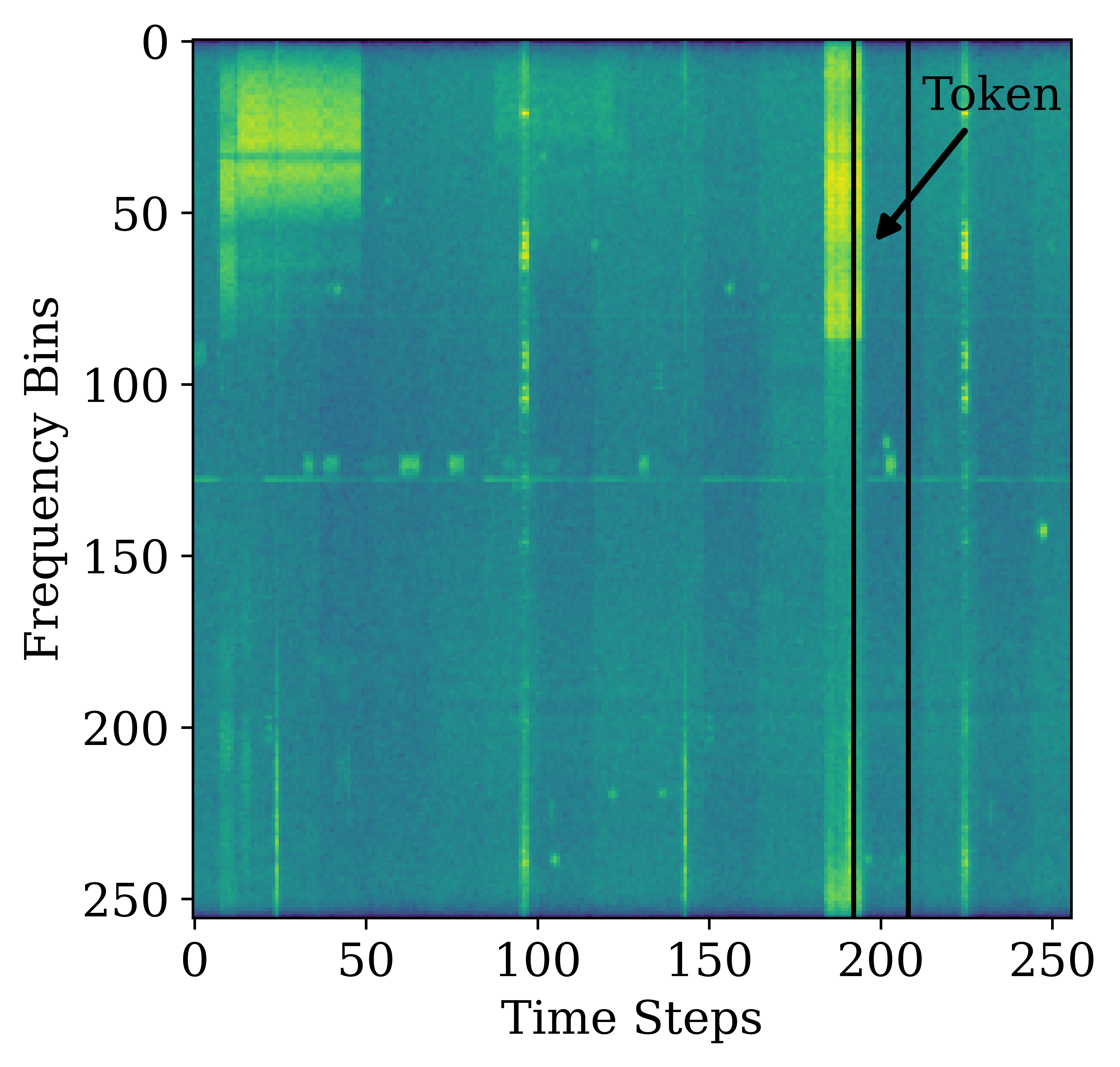}
    \caption{A radio sentence created from the RRD dataset.}
    \label{fig:sentence}
\end{figure}

A sample sentence is illustrated in Figure \ref{fig:sentence}, where the token is also labelled. Next we discuss our proposed approach to use these radio sentences and tokens to pretrain and develop a radio foundation model. 

\subsection{Masked Spectrogram Modelling}

\begin{figure*}
    \centering
    \includegraphics[width=0.67\textwidth, keepaspectratio]{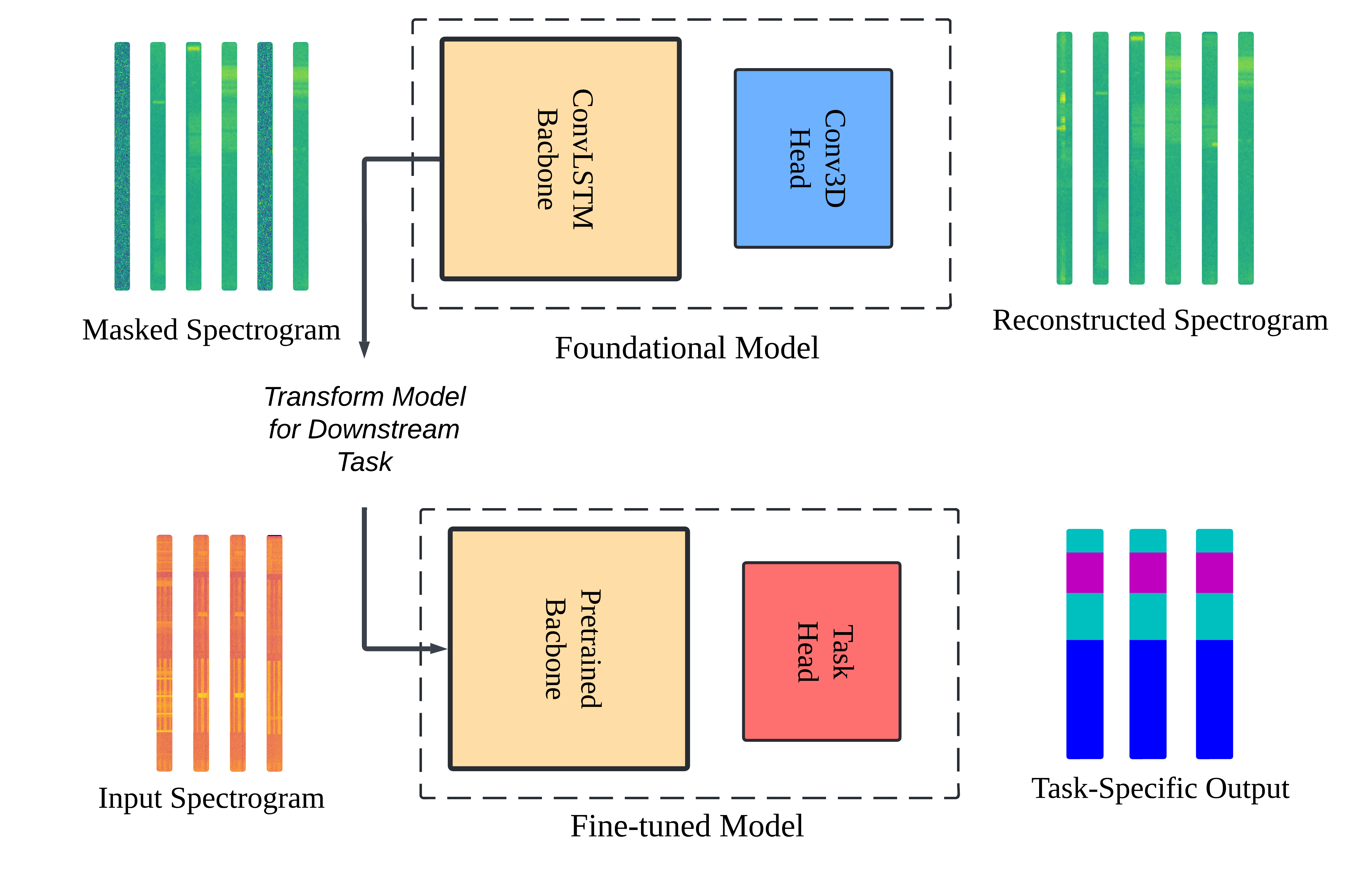}
    \caption{Illustration of the proposed methodology for MSM pretraining and downstream task fine-tuning.}
    \label{fig:msm_foundational_model}
\end{figure*}

\begin{figure}[h!]
    \centering
    \includegraphics[width=0.9\linewidth, keepaspectratio]{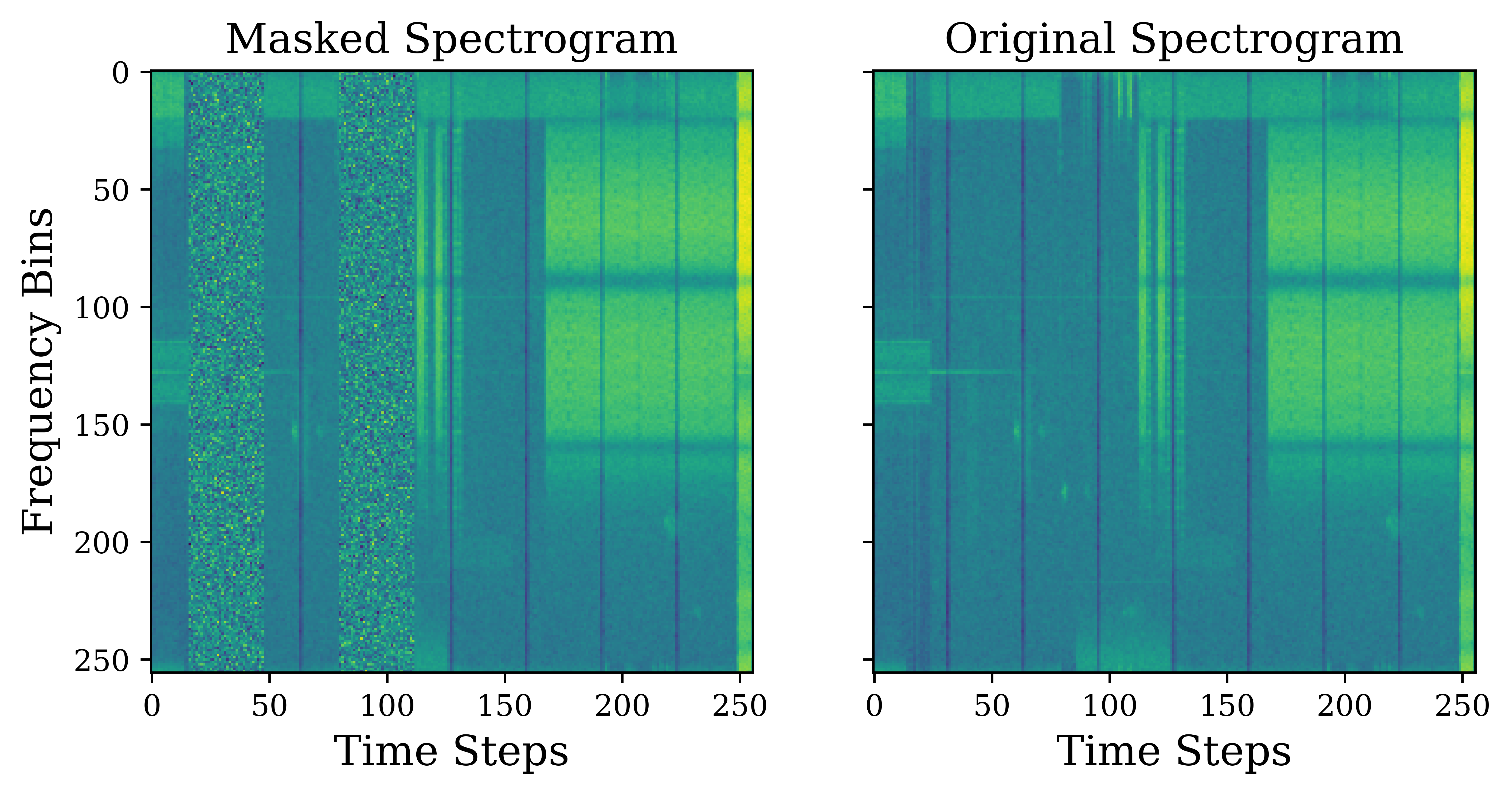}
    \caption{Masked and original spectrogram pair.}
    \label{fig:msm_sample}
    \vspace{-0.2cm}
\end{figure}

We propose a technique, we refer to as masked spectrogram modeling (MSM) for pretraining the foundational model. This approach involves inputting a spectrogram into a deep learning model and masking a portion of it—typically 20\%. Masking involves replacing the actual content of the spectrogram with white noise as shown in Figure \ref{fig:msm_sample}. The model's objective is to reconstruct the original spectrogram from the masked version, effectively denoising it in the process. To achieve this, the model analyzes the surrounding context and infers what was likely in the masked positions. Throughout the learning process, the model is expected to develop an understanding of radio signals as represented by spectrograms, creating an internal representation that enables it to accurately recover the original spectrograms. A notable advantage of this approach is that it operates without the need for labels. Radio signals can be recorded and fed directly into the model pipeline, which then leverages them to refine its internal representation. The pretrained model can then be fine-tuned for any related downstream task, the procedure is illustrated in Figure \ref{fig:msm_foundational_model} and a general algorithm is described in Algorithm \ref{Alg:msm}.

\begin{table}[h!]      
\renewcommand{\arraystretch}{1.5}
\caption{ConvLSTM Hyperparameters}
\setlength{\tabcolsep}{12pt}
\centering
\begin{tabular}{|c|c|}
\hline
\textbf{Parameter}          & \textbf{Value}                     \\ \hline
Layers                      & 5 ConvLSTM + 1 Conv3D                \\ \hline
Number of kernels per layer & 64                                 \\ \hline
Kernel size                 & 3 \\ \hline
Activation function         & ReLU                             \\ \hline
\end{tabular}
\vspace{5pt}

\label{tab:msm_model_params}
\end{table}

\begin{algorithm}
\DontPrintSemicolon
\SetAlgoLined
\SetKwInOut{Input}{Input}
\SetKwInOut{Output}{Output}
\textbf{Pretraining subalgorithm}\;
\Input{\textit{initial\_model}, \textit{IQ\_recordings}}
\Output{\textit{pretrained\_model}}
\textit{pretrained\_model} $\gets$ \textit{initial\_model}\;
Convert \textit{IQ\_recordings} to radio sentence representation using the procedure described in Section \ref{sec:datasets}\;
\For{\normalfont\textit{sentence} in all \textit{radio\_sentences}}{
\textit{masked\_sentence} $\gets$ RANDOM\_MASK(\textit{sentence})\;
\textit{predicted\_sentence} $\gets$ \textit{pretrained\_model}$\cdot$FORWARD(\textit{masked\_sentence})\;
\textit{loss} $\leftarrow$ $\text{MSE}_{\text{MSM}}$(\textit{masked\_sentence}, \textit{sentence}) using eq. \eqref{eq:loss_masked}\;
\textit{pretrained\_model} $\gets$ UPDATE(\textit{pretrained\_model}, \textit{loss})\;
}
\textbf{end subalgorithm}\;
\vspace{5pt}
\hrule
\vspace{5pt}
\textbf{Fine-tuning subalgorithm}\;
\Input{\textit{pretrained\_model}, \textit{downstream\_dataset}}
\Output{\textit{finetuned\_model}}
\textit{finetuned\_model} $\gets$ \textit{pretrained\_model}\;
\textit{preprocessed\_dataset} $\gets$ PREPROCESS(\textit{downstream\_dataset}) using the same transformations utilized for the \textit{pretrained\_model}\;
\For{\normalfont every (\textit{input, target}) in \textit{preprocessed\_dataset}}{
\textit{prediction} $\gets$ \textit{finetuned\_model}$\cdot$FORWARD(\textit{input})\;
\textit{loss} $\gets$ LOSS\_FN(\textit{prediction}, \textit{target})\;
\textit{finetuned\_model} $\gets$ UPDATE(\textit{pretrained\_model}, \textit{loss})\;
}
\textbf{end subalgorithm}\;
\caption{Self-Supervised Pre-training and Downstream Fine-Tuning Framework}
\label{Alg:msm} 
\end{algorithm}

A ConvLSTM model is used for pretraining, utilizing the hyperparameters listed in Table \ref{tab:msm_model_params}. The mean-square error is used as the loss function, computed only for the \emph{masked tokens}. The loss function $\mathcal{L}_{\text{MSM}}$ of MSM task can be written as:
\begin{equation}
    \label{eq:loss_masked}
    \mathcal{L}_{\text{MSM}} = \sum_{n=1}^{N} \sum_{t=1}^{T}  \left\|\myvec\left(\mathbf{W}_t^{(n)}\right) - \myvec\left(\hat{\mathbf{W}}_t^{(n)}\right)\right\|_2^2 \mathbb{I}_{\text{masked}}(n, t) 
\end{equation}
where $N$ is the batch size, $\mathbf{W}_t^{(i)} \in \mathbb{R}^{256 \times 16}$ is the $t^{\text{th}}$ input token from sample $n$, $\hat{\mathbf{W}}_t^{(i)} \in \mathbb{R}^{256 \times 16}$ is the  $t^{\text{th}}$ predicted token for sample $n$, $T$ is the number of input tokens, $\myvec$ denotes the vectorization operation, $\|\cdot\|_2$ is the $L_2$ norm and $\mathbb{I}_{\text{masked}}(i, t)$ is an indicator function that outputs $1$ if token $t$ from sample $n$ was masked and $0$ otherwise. The resulting self-supervised pre-trained model serves as our radio foundational model and is used for the following two downstream radio tasks.

\subsection{Spectrum Forecasting}
In this task, the model takes a sequence of tokens $\left\{\mathbf{W}_1^{(n)}, \cdots, \mathbf{W}_T^{(n)}\right\}$ as input and aims to predict the next token $\hat{\mathbf{W}}_{T+1}^{(n)}$. Training involves minimizing the mean-square-error loss between the predicted token and the actual next token for a batch of inputs $\left(\left\{\mathbf{W}_1^{(n)}, \cdots, \mathbf{W}_T^{(n)}\right\}_{n=1}^{N}, \left\{\mathbf{W}_{T+1}^{(n)}\right\}_{n=1}^{N}\right)$, where $n$ is the sample index and $N$ represents the batch size. This loss function, denoted as $\mathcal{L}_{\text{SF}}$, is formulated as follows:
\begin{equation}
    \label{eq:loss_forecasting}
    \mathcal{L}_{\text{SF}} = \sum_{n=1}^{N} \left\|\myvec\left(\mathbf{W}_{T+1}^{(n)}\right) -  \myvec\left(\hat{\mathbf{W}}_{T+1}^{(n)}\right)\right\|_2^2
\end{equation}
The model architecture described in Table \ref{tab:msm_model_params} is used, yet with two notable modifications: first, only the initial token of the output Conv3D layer is considered, disregarding the remaining outputs. Second, the backbone of the model, consisting of the 5 ConvLSTM layers, is frozen, and its weights are initialized with those obtained from the MSM task. Consequently, the features learned during the MSM task are utilized, while only the final layer undergoes fine-tuning for spectrum forecasting purposes. 

\subsection{Spectrogram Segmentation}
In this task, the model processes a spectrogram of size $(256, 256)$, which is then tokenized into 16 tokens of shape $(256, 16)$. Consequently, the input comprises a sequence of tokens $\left\{\mathbf{W}_1^{(n)}, \cdots, \mathbf{W}_{16}^{(n)}\right\}$ while the output is a segmented image $\mathbf{Y}^{(n)} \in \{0, 1\}^{(256,\ 256,\ 3)}$ that is one-hot encoded and the model prediction is denoted as $\hat{\mathbf{Y}}^{(n)} \in [0, 1]^{(256,\ 256,\ 3)}$. For a batch of size $N$, we utilize the cross entropy loss written as:
\begin{equation}
    \label{eq:loss_segmentation}
    \mathcal{L}_{\text{SG}} = -\sum_i \sum_j \sum_{n=1}^{N} \mathbf{Y}_{ij}^{(n)} \cdot \log\left(\hat{\mathbf{Y}}_{ij}^{(n)}\right)
 \end{equation}
Similar to spectrum forecasting, the backbone consists of 5 ConvLSTM layers. Subsequently, the backbone's output is concatenated to form a shape of $(256, 256)$, serving as input to a two-layer Conv2D classifier. The backbone's weights remain fixed, the classifier is fine-tuned for the segmentation task.
\section{Results and Discussion}
\label{sec:results}

This section evaluates the proposed self-supervised radio pre-training methodology by comparing the performance of the resulting foundational model when fine-tuned on two downstream tasks—spectrum forecasting and segmentation—to a baseline. The baseline model shares the same architecture but is trained from scratch on identical data. 

\subsection{Downstream Task-1: Spectrum Forecasting}

\textit{Data and Model Training}: We partition the RRD dataset, allocating 50\% for pretraining and reserving the remaining 50\% for forecasting. 

We fine-tune the pretrained model using the remaining 50\% of the RDD dataset, which we further split into training and test sets with an 80\% to 20\% ratio.

\begin{figure}[t!]
    \centering
    \includegraphics[width=\linewidth, keepaspectratio]{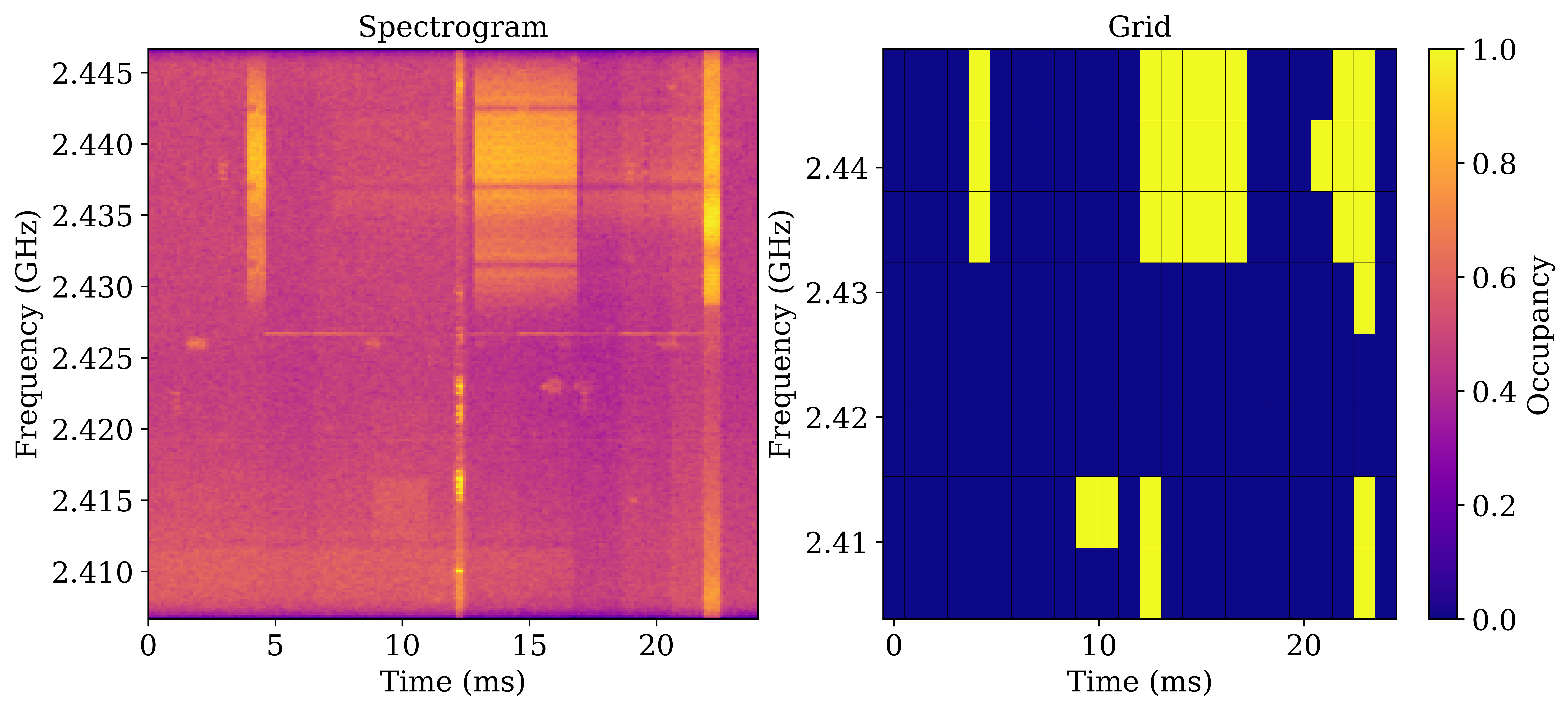}
    \caption{Illustration of a spectrogram and its corresponding resource grid for a block size of ($1$ ms, $5$ MHz).}
    \label{fig:spect2grid}
\end{figure}

\textit{Evaluation Metric}: To evaluate the model's forecasting performance, relying solely on visual comparison between the target and predicted spectrograms will not suffice. Therefore, we adopt a more robust metric by transforming each spectrogram into a resource grid composed of resource blocks (RBs) with predefined time and frequency resolutions. This process involves dividing the spectrogram into these blocks and computing the mean value for each. Subsequently, a threshold is applied to the resulting grid, rendering it binary—where a value of $1$ denotes an occupied block and $0$ denotes a vacant one. Figure \ref{fig:spect2grid} depicts the outcome of this transformation. The threshold $\delta$ is empirically determined as:
\begin{equation}
\label{eq:threshold}
\delta = \mu + 0.5 \times \sigma
\end{equation}
where $\mu$ and $\sigma$ represent the mean and standard deviation of the spectrogram, respectively.

Our primary evaluation metric focuses on the model's capacity to correctly predict the occupancy of a resource block when it is indeed occupied. Predicting vacancy is straightforward, given its prevalence as the dominant class. From the perspective of opportunistic spectrum access, it is crucial to consistently detect occupied blocks to mitigate potential collisions. 

\begin{figure}[t!]
    \centering
    \includegraphics[width=0.9\linewidth, keepaspectratio]{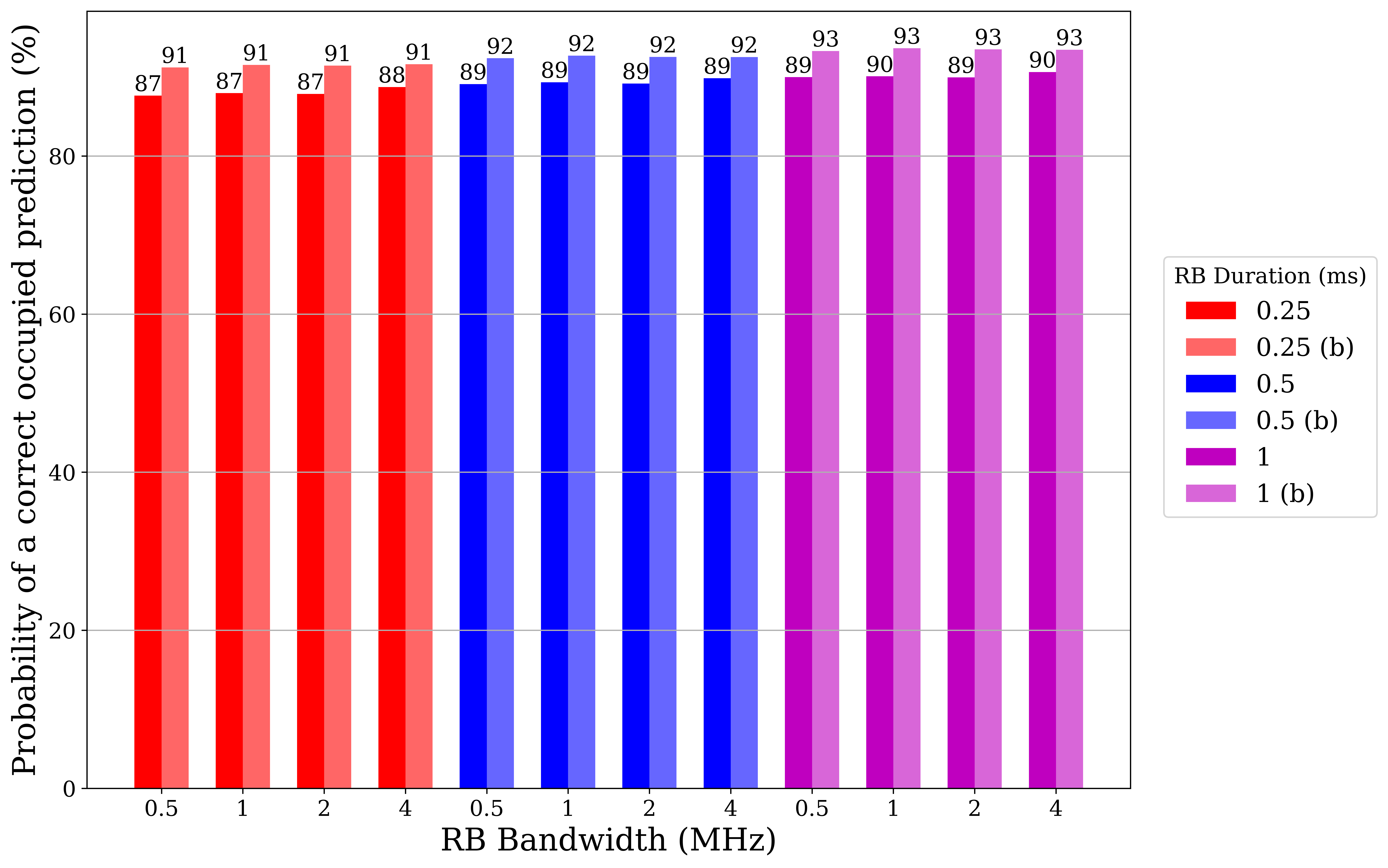}
    \caption{Probability of correct occupied predictions. The solid lines are the foundational tuned model and (b) is the baseline.}
    \label{fig:p_occupied}
\end{figure}
This metric is depicted in Figure \ref{fig:p_occupied} for predictions extending 4 tokens into the future, across various time and frequency resolutions for the resource block, with (b) representing the baseline. The specialized baseline outperforms the tuned foundational model, though by a small margin.

\subsection{Downstream Task-2: NR-LTE Segmentation}
For the segmentation task, we utilize the SD dataset to fine-tune the foundational model. The main challenge here lies in the distinct nature of the spectrograms within this dataset compared to those used during the self-supervised pre-training. Consequently, the learned features may not generalize as effectively to segmentation as they do to forecasting. In addition, segmentation is a classification task and the model was pre-trained on regression only. Our objective is to examine the extent to which the learned representations of the foundational model can generalize under such distinct data distributions and across task types. 

\begin{figure}[!t]
    \centering
    \begin{minipage}[b]{0.45\linewidth}
        \centering
        \includegraphics[width=\linewidth, keepaspectratio]{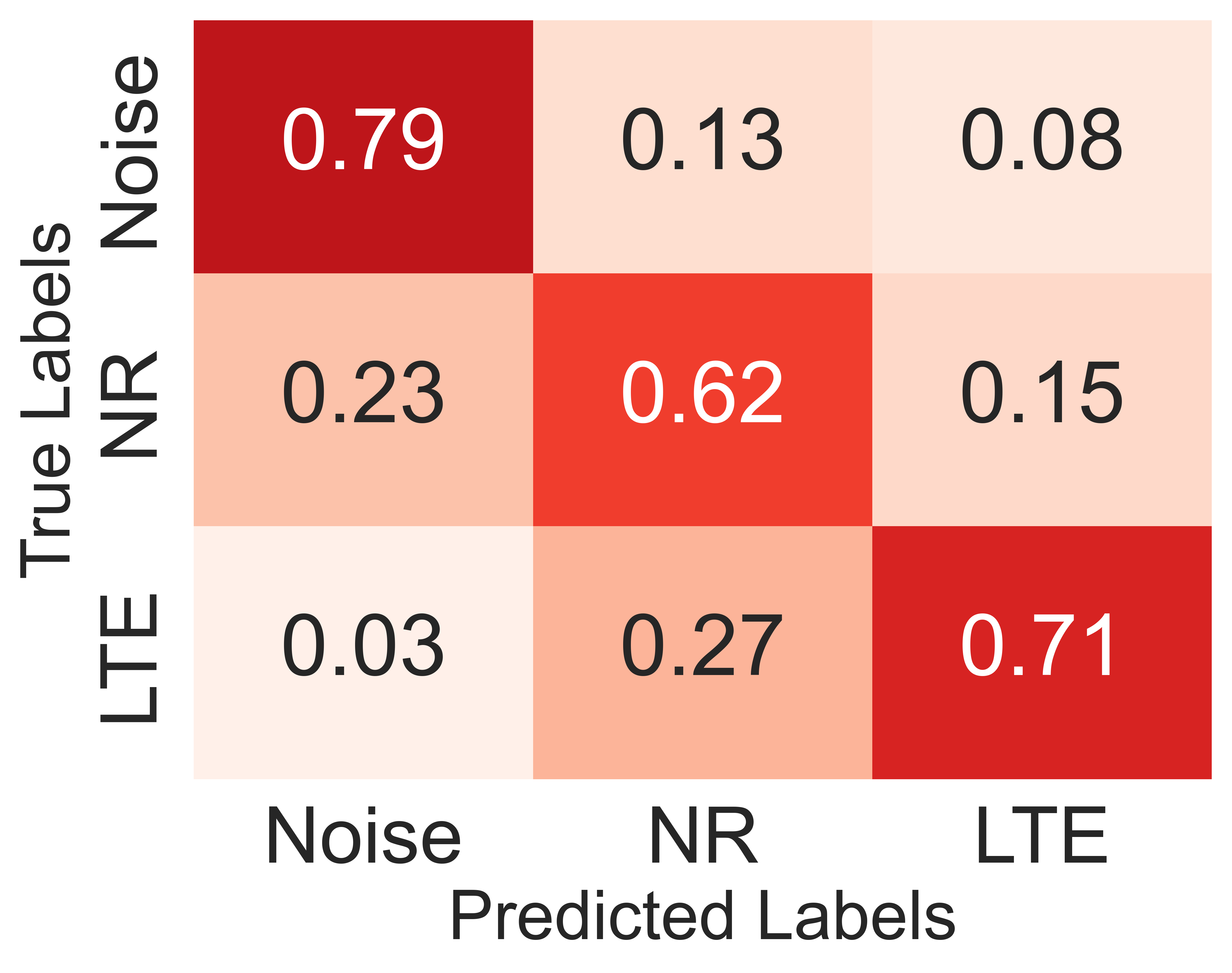}
        \caption*{Tuned Foundational Model}
        \label{fig:model_conf_mat}
    \end{minipage}
    \hfill
    \begin{minipage}[b]{0.45\linewidth}
        \centering
        \includegraphics[width=\linewidth, keepaspectratio]{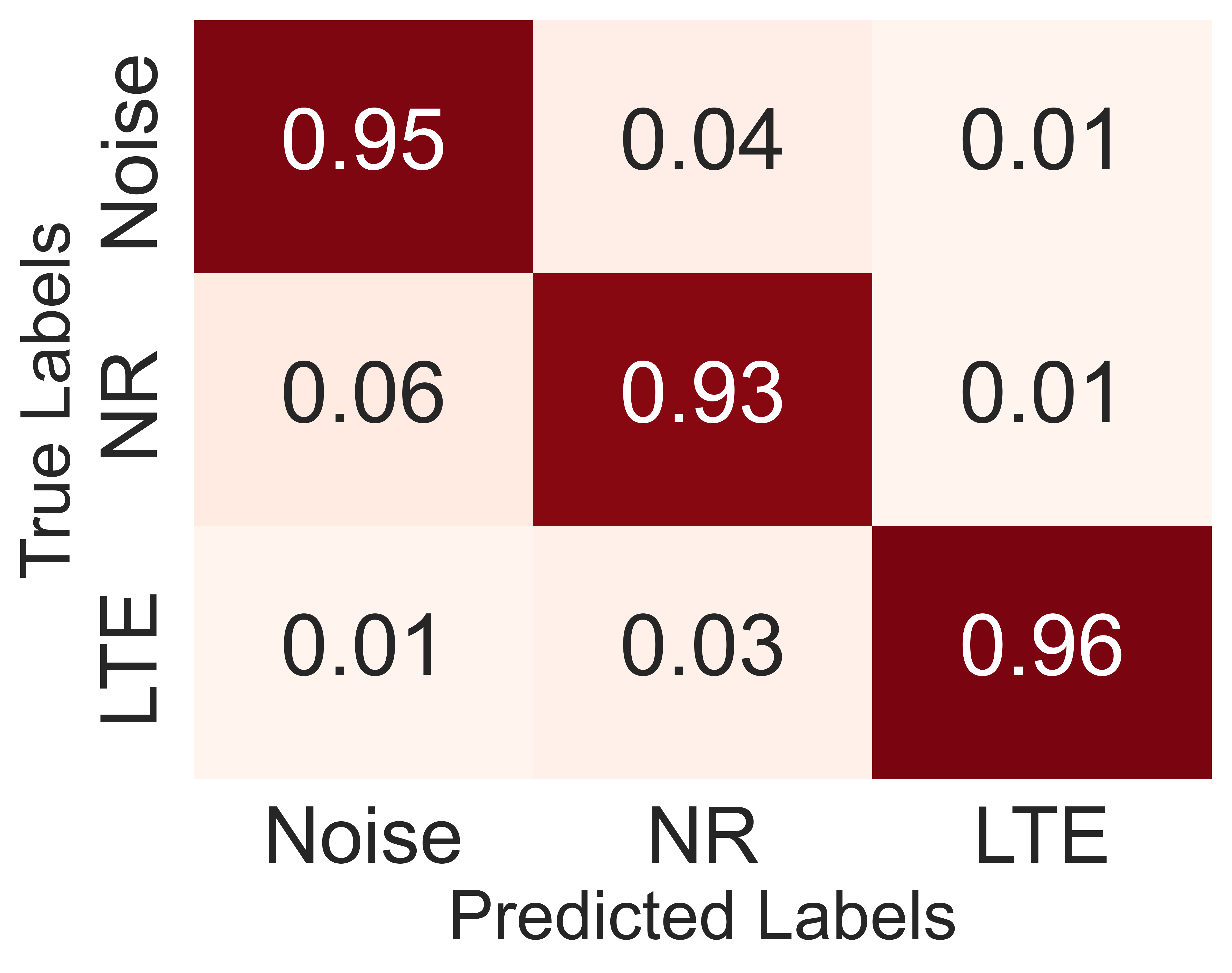}
        \caption*{Baseline Model}
        \label{fig:baseline_confmat}
    \end{minipage}
    \caption{Segmentation performance using confusion matrices.}
    \label{fig:seg_conf_mat}
\end{figure}

We evaluate the model's performance using confusion matrices for the three classes: Noise, NR, and LTE, which quantify the model's prediction accuracy for each class. Figure \ref{fig:seg_conf_mat} presents the confusion matrices for both the baseline and fine-tuned models. Notably, the fine-tuned model struggles with distinguishing NR signals, while the baseline model demonstrates strong performance across all classes. This suggests that the features provided by the pretrained backbone are not sufficiently discriminative to differentiate NR signals from other classes, though they perform adequately in separating signals (NR or LTE) from noise. A more complex head and fine-tuning training process may also be needed.

\begin{figure}[!t]
    \centering
    \begin{minipage}[b]{0.44\linewidth}
        \centering
        \includegraphics[width=\linewidth, keepaspectratio]{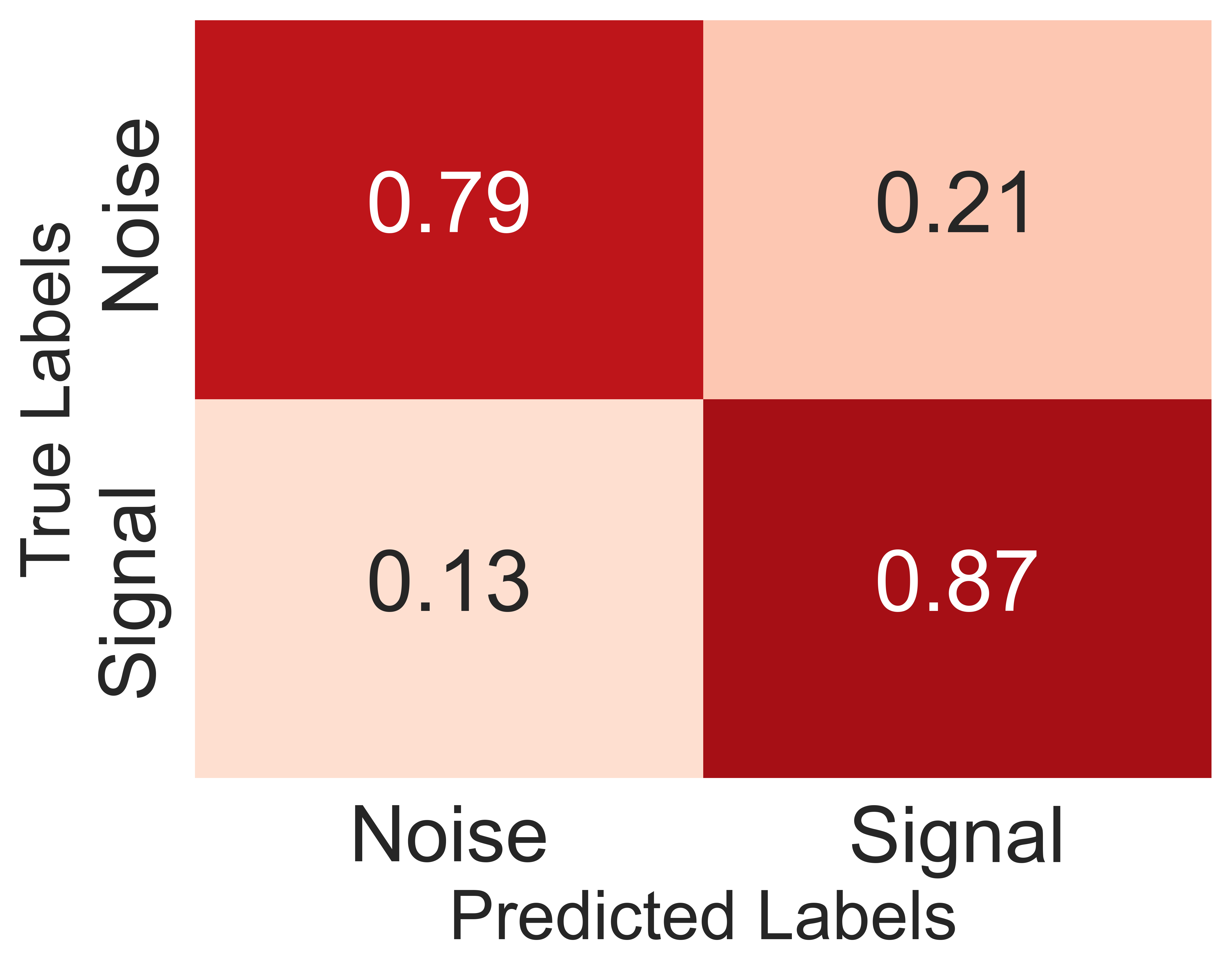}
        \caption*{Tuned Foundational Model}
        \label{fig:binary_model_confmat}
    \end{minipage}
    \hfill
    \begin{minipage}[b]{0.54\linewidth}
        \centering
        \includegraphics[width=\linewidth, keepaspectratio]{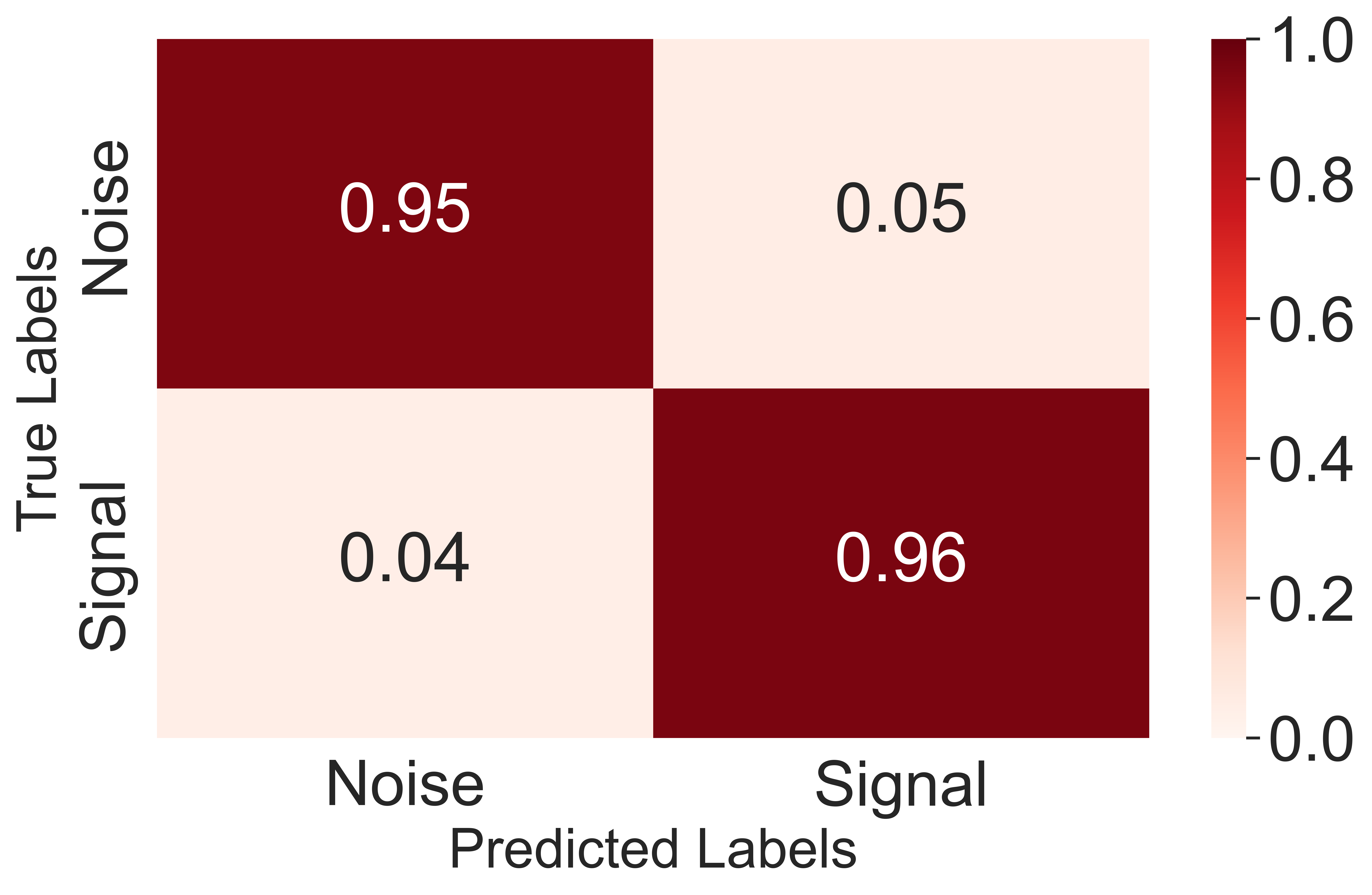}
        \caption*{Baseline Model}
        \label{fig:binary_baseline_confmat}
    \end{minipage}
    \caption{Binary segmentation performance using confusion matrices.}
    \label{fig:binary_seg_conf_mat}
\end{figure}

To further illustrate this, we simplify the task to binary segmentation, merging NR and LTE into a single \textit{signal} class. The resulting confusion matrices are shown in Figure \ref{fig:binary_seg_conf_mat}. Here, while the baseline model still outperforms the fine-tuned model, the correct detection of signals is higher. We attribute this to differences in data distribution between the pretraining and SD datasets. Pretraining on a larger and more diverse dataset may help bridge this gap. Further research by the community will be needed in these directions to build large-scale foundational radio models.

\section{Conclusion}
\label{sec:conclusion}

In this paper, we introduced a self-supervised radio pretraining approach, MSM, to build a foundational model for spectrogram learning. Drawing inspiration from the success of foundational DL models in various domains, the goal was to learn features that could generalize to related downstream tasks. We demonstrated that by fine-tuning the developed MSM model for two downstream tasks: spectrum forecasting and segmentation. 
Our results show that the fine-tuned models exhibited competitive performance compared to baselines trained from scratch, while requiring much less training time to converge. We believe that extending the proposed MSM approach to larger models and utilizing large-scale, diverse datasets for pretraining has the potential to develop robust radio foundational models that yield competitive performance across various spectrogram learning tasks.

\bibliography{bibliography.bib}
\bibliographystyle{IEEEtran}

\end{document}